\documentclass[%
 amsmath,amssymb,superscriptaddress,
 reprint,%
 showkeys,
 onecolumn,
]{revtex4-2}

\usepackage[utf8]{inputenc}
\usepackage[T1]{fontenc}
\usepackage{float}
\usepackage{dcolumn}
\usepackage{longtable}
\usepackage{subfigure}

\usepackage[dvipsnames]{xcolor}

\usepackage{soul}

\usepackage{multirow}
\usepackage[subpreambles=true]{standalone}
\begin{document}

\preprint{AIP/123-QED}

\title{Commissioning of the novel Continuous Angle Multi-Energy Analysis Spectrometer at the Paul Scherrer Institut}

\author{Jakob Lass}
\affiliation{Laboratory for Neutron Scattering and Imaging, Paul Scherrer Institut, CH-5232 Villigen PSI, Switzerland}
\affiliation{Nanoscience Center, Niels Bohr Institute, University of Copenhagen, DK-2100 Copenhagen Ø , Denmark}
\author{Henrik Jacobsen}
\affiliation{Laboratory for Neutron Scattering and Imaging, Paul Scherrer Institut, CH-5232 Villigen PSI, Switzerland}
\affiliation{Nanoscience Center, Niels Bohr Institute, University of Copenhagen, DK-2100 Copenhagen Ø , Denmark}
\author{Kristine M. L. Krighaar}
\affiliation{Nanoscience Center, Niels Bohr Institute, University of Copenhagen, DK-2100 Copenhagen Ø , Denmark}
\author{Dieter Graf}
\affiliation{Laboratory for Neutron and Muon Instrumentation, Paul Scherrer Institute, CH-5232 Villigen PSI, Switzerland}
\author{Felix Groitl}
\affiliation{Laboratory for Neutron Scattering and Imaging, Paul Scherrer Institut, CH-5232 Villigen PSI, Switzerland}
\affiliation{Laboratory for Quantum Magnetism, Ecole Polytechnique Fédérale de Lausanne (EPFL), CH-1015 Lausanne, Switzerland}
\author{Frank Herzog}
\author{Masako Yamada}
\author{Christian Kägi}
\author{Raphael Müller}
\author{Roman Bürge}
\author{Marcel Schild}
\author{Manuel S. Lehmann}
\author{Alex Bollhalder}
\author{Peter Keller}
\author{Marek Bartkowiak}
\author{Uwe Filges}
\author{Urs Greuter}
\author{Gerd Theidel}
\affiliation{Laboratory for Particle Physics, Paul Scherrer Institute, CH-5232 Villigen PSI, Switzerland}
\author{Henrik M. Rønnow}
\affiliation{Laboratory for Quantum Magnetism, Ecole Polytechnique Fédérale de Lausanne (EPFL), CH-1015 Lausanne, Switzerland}
\author{Christof Niedermayer}
\author{Daniel G. Mazzone}
\email{The authors to whom correspondence may be addressed: jakob.lass@psi.ch, daniel.mazzone@psi.ch}
\affiliation{Laboratory for Neutron Scattering and Imaging, Paul Scherrer Institut, CH-5232 Villigen PSI, Switzerland}

\date{\today}% It is always \today, today,
             %  but any date may be explicitly specified
\begin{abstract}
We report on the commissioning results of the cold neutron multiplexing secondary spectrometer CAMEA (\textbf{C}ontinuous \textbf{A}ngle \textbf{M}ulti-\textbf{E}nergy \textbf{A}nalysis) at the Swiss Spallation Neutron Source (SINQ) at the Paul Scherrer Institut, Switzerland. CAMEA is optimized for an efficient data acquisition of scattered neutrons in the horizontal scattering plane, allowing for detailed and rapid mapping of low-energy excitations under extreme sample environment conditions. 
\end{abstract}


\maketitle

\section{Introduction}
Modern neutron spectrometers enable a detailed mapping of elemental excitations in solid state materials, which provide fundamental insight into the microscopic interactions among lattice and electronic degrees of freedom. Almost all neutron scattering experiments suffer from the fact that the technique is intrinsically flux limited, which has led to continuous efforts in improving its efficiency. Notably, this consists of increasing the number of neutrons that are created at the source or arrive at the sample position, and finding refined solutions to efficiently determine the energy of the scattered neutrons and detect them over large solid angles. % while covering a large part of reciprocal space.
In fact, obtaining a large detector coverage is one of the major improvements that came with the invention of direct geometry time-of-flight (ToF) spectrometers. Here, a time-resolved white beam of neutrons impinges on the sample and is detected in large detector banks spanning a large solid angle. The final energy of the scattered neutrons is deduced by the time the neutrons take to reach the detector. The time-stamping of individual neutrons is achieved by pulsed neutron sources or/and through complex chopper systems that substantially reduce the neutron flux. Recent examples of modern ToF spectrometers include Panther at the Institut Laue-Langevin~\cite{Deen2010}, CNCS at Oak Ridge National Laboratory~\cite{Ehlers2011}, LET at the ISIS Muon and Neutron Source~\cite{Bewley2011}, AMATERAS at J-PARC~\cite{Nakajima2011}, and C-SPEC which is currently under construction at the European Spallation Source~\cite{Deen2021}.

An alternative option to increase the angular detector coverage of neutron spectrometers are monochromatic multiplexing spectrometers. One main advantage of these instruments is that the incident neutron beam does not need to have a time structure, thus profiting from the large neutron flux of triple-axis instruments. The incoming energy is selected by a monochromator and the final energies are determined via analyzer crystals that are spread over a large solid angle. Since the monochromator is rotated as a function of the initial energy, the size and weight of the instrument is limited. This is because the secondary spectrometer consists of large detector and/or analyzer banks that have to physically move during an experiment. Examples of monochromatic multiplexing spectrometers that are currently in operation or under construction are the MACS spectrometer at NIST~\cite{Rodriguez2008}, FlatCone at the Institut Laue-Langevin~\cite{Kempa2006}, BAMBUS, PUMA, and MultiFLEXX at FRM-II~ \cite{Sobolev2015,Lim2014,Groitl2017MultiFLEXX}, BIFROST at the European Spallation Source~\cite{Bifrost} and the Continuous Angle Multi-Energy Analysis (CAMEA) at the Paul Scherrer Institut (PSI) ~\cite{Groitl2016} which replaced the RITA II spectrometer\cite{Lefmann2000}. CAMEA represents a new generation of multiplexing neutron spectrometers employing the prismatic analyzer concept which enables a quasi-continuous energy coverage while relying on a finite set of fixed final energies~\cite{Birk2014}. Here, we report on the commissioning of the CAMEA secondary spectrometer, including its neutron guide, and double-focusing monochromator. Details concerning the spectrometer design and the data reduction software MJOLNIR - dedicated to treat multiplexing data - are reported in Refs. ~\cite{Groitl2016,Lass2020}.

\subsection{Spectrometer and guide design}

\begin{figure}[ht!]
    \centering
    \subfigure[]{\includegraphics[width=0.8\linewidth]{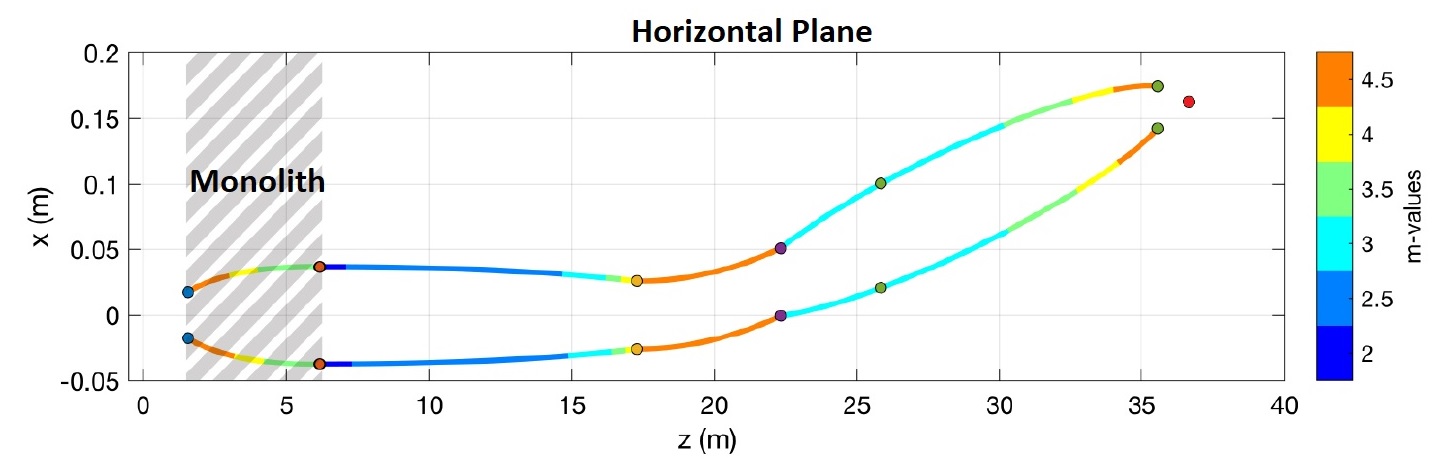}}
    \subfigure[]{\includegraphics[width=0.8\linewidth]{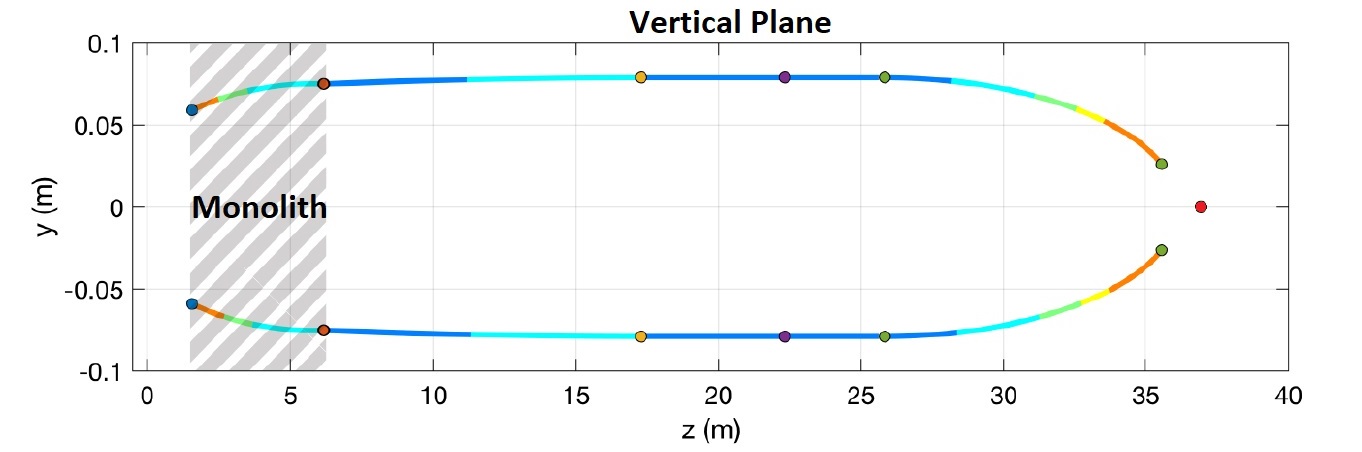}}
    \caption{Schematic overview of the CAMEA guide showing the employed m-values (represented in the colour scale) from the monolith to the virtual source (red point) that is located 1.2 m upstream of the monochromator (Reprinted from "Neutron guides RNR12 and RNR13"\cite{GuideUpgrade}) shown in (a) the horizontal and (b) vertical planes. The extraction ellipse is located closest to the source at z = 0 m and extends to a distance of 6 m. A horizontal focusing elliptical section extends from the monolith to the start of the bending section signified by the yellow dots. After the bend section the last expansion and focusing ellipse is implemented, focusing the beam to the virtual source. From the monolith to half-way through the last horizontal elliptical guide the vertical guide is almost constant. Only a focusing ellipse is present in the last 11 m, between the green dots.}
    \label{fig:guide}
\end{figure}

The neutron guide system of the CAMEA guide was upgraded during the extensive shutdown period of the SINQ in 2019-2020~\cite{GuideUpgrade}. The most noteworthy specifications of the guide is that it is optimized for neutrons in the wavelenght range from 2.5 to 5 \AA, accepting a large horizontal and vertical divergence of $\pm$ 2$^\circ$. This is achieved through a double elliptical guide profile with $m$-values of up to 4.5 (see Fig.~\ref{fig:guide}), where $m$ refers to the critical scattering angle through $m \lambda 0.1$ in $^\circ/$\AA\ for a neutron wavelength $\lambda$. In the horizontal plane, the guide is built from two ellipses separated by a bent section while vertically the guide has an elliptically expanding section followed by a straight part which finally, through an ellipse, focuses the beam into a virtual source. Space limitations required that the first elliptical guide segment after the moderator features a length of 1.5 m with a maximal opening of 3.5 x 14 cm$^2$. A bent second guide segment inside the bunker prevents a direct line of sight between moderator and monochromator, which substantially reduces parasitic scattering at the sample position. The second elliptical section was designed to focus the neutrons onto a virtual source located upstream of the monochromator.

The final design was optimized using Guide\_bot~\cite{GuideBot} against the performance of the guide for its brilliance at the sample position. Gold foil measurements at the end of the guide, showed that the guide upgrade resulted in a flux gain of about a factor of 5.5 compared to the RITA II guide due to a more divergent beam, which is consistent with our simulations\cite{JonasInternalReport}.

\begin{figure}[ht!]
    \centering
    \subfigure[]{\includegraphics[trim={0cm 0cm 0cm 0cm},clip,width=0.46\linewidth]{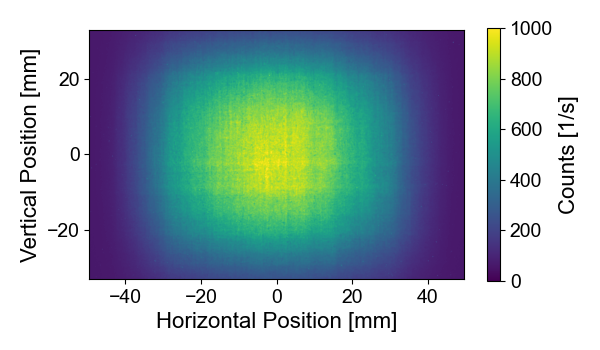}}
    \subfigure[]{\includegraphics[trim={0cm 0cm 0cm 0cm},clip,width=0.46\linewidth]{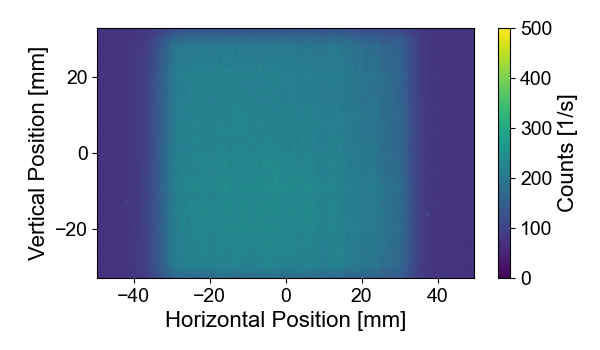}}
    \caption{Measured beam profile at the sample position at (a)  $E_i = $ 4.6 meV using an optimally focused monochromator and (b) $E_i = $ 5 meV for a flat monochromator.}
    \label{fig:CameraSample}
\end{figure}

The divergence of the neutron beam at the virtual source is optimized for a large, double focusing monochromator. The CAMEA monochromator consists of 187 individual 15 x 15 x 2 mm$^3$ highly ordered pyrolytic graphite (HOPG) single crystals with a mosaicity of 0.6$^\circ$ to 0.8$^\circ$. They are arranged in 11 rows and 17 columns, respectively, spanning a total size of 173 x 262 mm$^2$ (including the gaps between the individual crystals). The crystals were mounted on 2 mm thick B$_4$C pieces to reduce parasitic background contributions. The mechanical design of the monochromator allows to focus the beam in vertical (radius of curvature of $>$ 0.6 m) and horizontal (radius of curvature $>$ 1.3 m) direction by a rotation of the individual graphite crystals. An optimally focused monochromator results in a beam spot of 23 x 32 mm$^2$ (full-width at half maximum (FWHM) intensity) at the sample position, see Fig.~\ref{fig:CameraSample}. The flux increase compared to a flat monochromator is 2.55 and 1.5 when the monochromator is focused vertically and horizontally, respectively. 

%% C:\Users\lass_j\Documents\CAMEA2022\CameraInGuide\VideoCamera.py
\begin{figure}[ht!]
    \centering
    \includegraphics[width=0.6\linewidth]{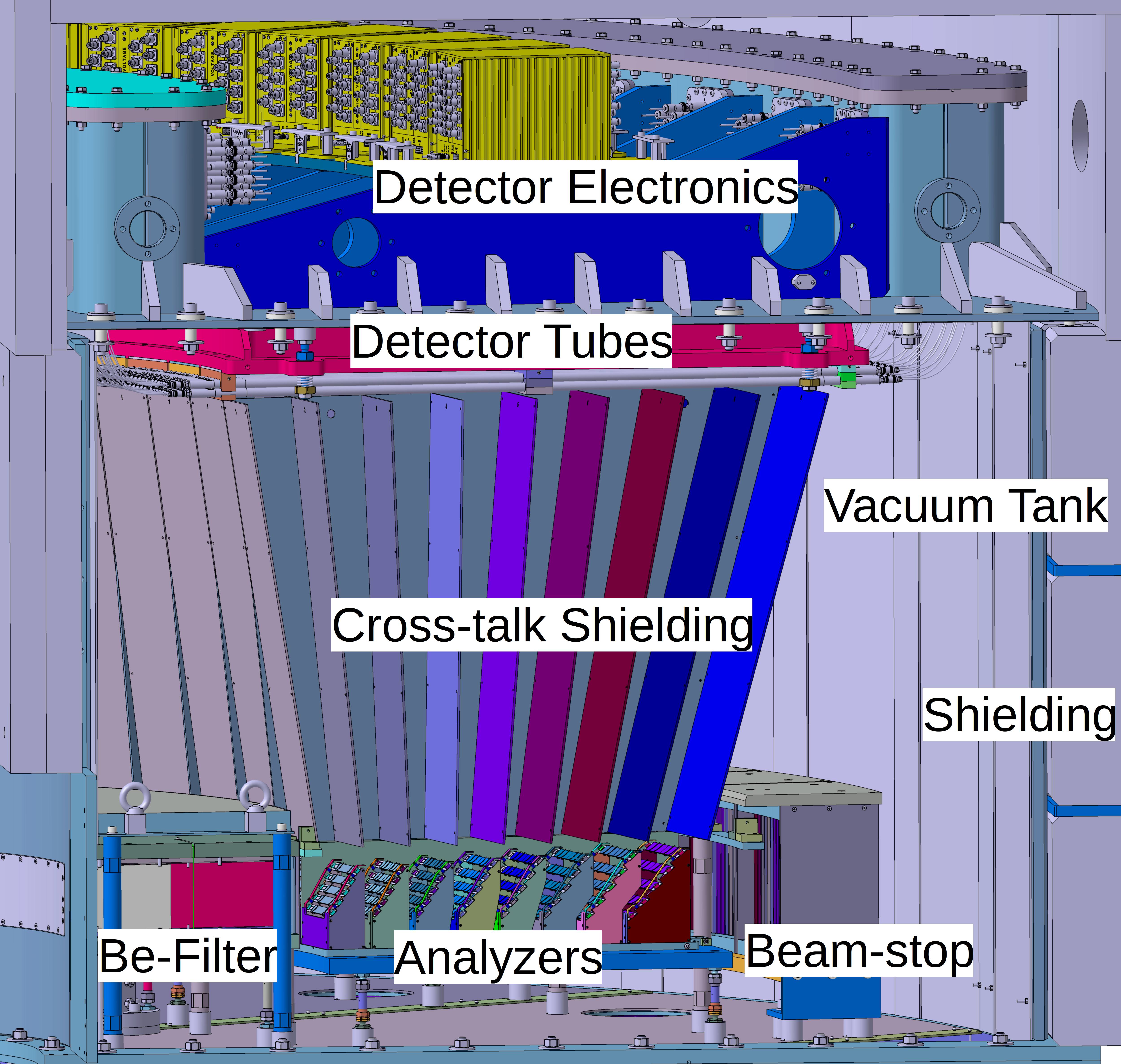}
    \caption{Side view of the CAMEA secondary spectrometer.}
    \label{fig:CAMEAsideview}
\end{figure}

The sample position is located 1.6 m downstream of the monochromator. Here neutrons are scattered into the secondary spectrometer, which is built in a modular design consisting of eight almost identical sections mounted next to one another (azimuthal direction) and separated by cross-talk shielding made of Boralcan. A side view of CAMEA showing one of the 8 sections is displayed in Fig.~\ref{fig:CAMEAsideview}. A detailed description of the design is reported in Ref.~\cite{Groitl2016}. The setup is optimized to detect neutrons scattered in the horizontal plane with an out-of-plane acceptance of $\pm$ 2$^{\circ}$. 

The instrument is compatible with extreme environments allowing the user to take advantage of the full sample environment suite at SINQ. The initial part of the secondary spectrometer's vacuum vessel ($p$ $<$ 4$\cdot$10$^{-7}$ mbar) consists of an Al window with a movable Boral slit and a combined radial collimator and Be-filter system~\cite{Groitl2016_collimator}. The filter-collimator combination is located at a distance of 57 cm from the sample position. The collimator is made of 28 cm long glass lamellas coated with 1 $\mu$m Gd and spaced by 1$^{\circ}$, efficiently cutting away any signal originating more than 8.6 cm away from the sample position. The first 12 cm between the lamellas consists of sintered Be wedges acting as an energy filter, providing a suppression of at least 10$^{5}$ of parasitic $\lambda/2$ contributions when cooled to around 100 K~\cite{Groitl2016_collimator,Groitl2016}. Such a suppression has been shown to be crucial for cold multiplexing instruments\cite{Toft-Petersen2016}.

After the collimator-filter system, the neutrons are scattered out of plane by one of eight HOPG analyzer segments which are set to scatter neutrons with final energies between 3.2 meV and 5.0 meV~\cite{Groitl2016}. Each angular segment hosts 13 1 m long position-sensitive 1/2 inch $^3$He tubes. They are arranged over two layers forming a W-shape pattern to cover the largest possible area above the analyzers with optimal angular resolution. A quasi-continuous energy coverage is achieved through a high mosaicity of the analyzer crystals in combination with distance collimation between the sample and analyzers as well as the analyzers and detectors~\cite{Birk2014}.
%The high mosaicity (1 degree) of the analyzer crystals combined with the distance collimation between sample position and analyzers, and between analyzers to detectors enables a quasi-continuous energy coverage~\cite{Birk2014}. 

The background of the open instrument geometry is reduced by a honeycomb cross-talk Boralcan shielding that is mounted between all analyzer segments. The entire vacuum vessel is made of non-magnetic steel and dressed by 20 cm thick neutron absorbing borated polyethylene blocks. The inside wall of the vacuum vessel is covered by Boralcan and also features a beam stop to further minimize background sources. The resulting background amounts to 0.3 counts per detector tube and minute during SINQ operation but with closed shutters. A low electronic background noise is achieved by the second generation detector electronics, also providing the opportunity for on-the-fly surveillance and adjustments of the gamma thresholds of each detector tube.

The engineering and construction phases of the CAMEA secondary spectrometer triggered the development of multiple in-house solutions on which we comment briefly in this paragraph. In order to mount all 600 analyzer crystals in a reproducible and efficient way, we used an off-cut silicon wafer and aluminium clips, which allowed us to apply enough force to prevent the graphite crystals from movements while still minimizing the background arising from the mounting material \cite{Groitl2017HOPG}. We developed special connection plugs that allowed us to operate the 104 detector tubes under high-voltage inside the evacuated steal tank. We also had to develop specialized air-cushions that can lift the 8.5 ton heavy instrument and position it with a precision of 0.1$^\circ$ at arbitrary angles on the instrument floor.

\section{Performance}

\begin{figure} % Made using C:\Users\lass_j\Documents\Software\Scripts\CountingTimes.py
    \centering
    \subfigure[]{\includegraphics[width=0.45\linewidth]{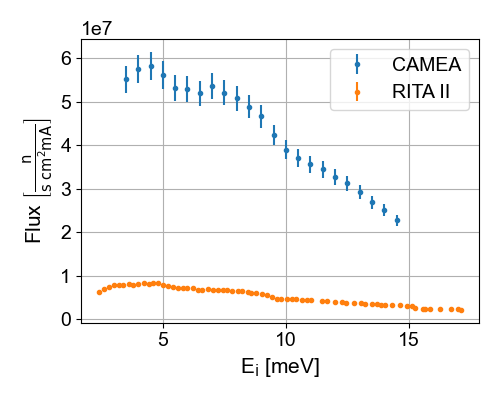}}
    \subfigure[]{\includegraphics[width=0.45\linewidth]{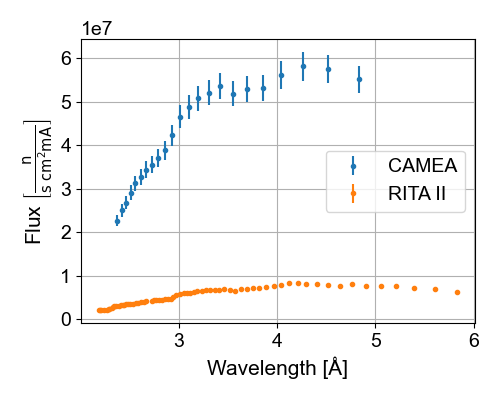}}
    \caption{Measured lethargy flux at the sample position (area of 10 x 10 mm$^2$) normalized to 1 mA on the SINQ target as function of (a) energy and (b) wavelength. The current performance of the primary spectrometer is compared to the replaced primary spectrometer RITA II.}
    \label{fig:Flux}
\end{figure}

The large phase-space transport of the double elliptical guide and double-focusing monochromator allows to tune the incident energy continuously between 3.2 meV and 15 meV with a maximal neutron intensity around 4.5 meV ($\lambda =$ 4.25 \AA), see Fig.~\ref{fig:Flux}. The angular coverage of the secondary spectrometer is $\pm 30.5^{\circ}$ relative to the centre of the detector tank which we denote as $2\theta$. In a standard experiment, a continuous angular coverage is achieved by a combination of two identical sample rotations scans where the detector tank is shifted by 4$^\circ$ in $2\theta$. This is needed to cover the blind spots arising from the wedges between the 8 analyzer modules (see Ref \cite{Groitl2016} for details). The resolution in $2\theta$ has been determined using a standard AlO sample of 1 cm in diameter at an incident energy $E_i$ = 5.0 meV. The FWHM of the resolution limited nuclear Bragg peaks were found to be 2.24(1)$^{\circ}$ for a vertically and horizontally focused monochromator. 

% Table generated using C:\Users\lass_j\Documents\Software\Alignment2022\BinningTable.py and normalization camea2022n000358
\begin{table}[H]
   \centering
    \begin{tabular}{|c|cccccccc|}\hline 
      Analyzer & 0 & 1 & 2 & 3 & 4 & 5 & 6 & 7\\\hline
      E$_f$ [meV] & 3.200 & 3.382 & 3.574 & 3.787 & 4.035 & 4.312 & 4.631 & 4.987\\
      FWHM [$\mu$eV] & 145 & 154 & 165 & 182 & 196 & 218 & 247 & 278\\\hline
    \end{tabular}
    \caption{Mean final energies and FWHMs for prismatic binning 1.}
    \label{tab:PrismaticVanadiumMeanBinning1}
\end{table}

\begin{figure}[ht!]
    \centering
    \includegraphics[width=0.45\linewidth]{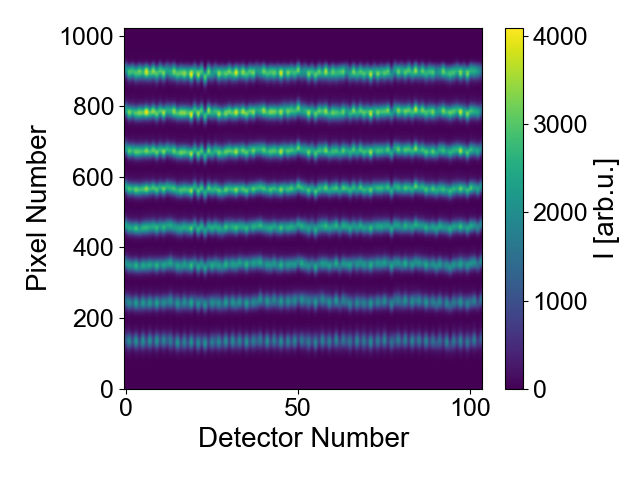}
    \caption{Neutron intensity from a standard vandium sample across all position sensitive detector tubes after integration over an incoming energy range between 2.9 meV and 5.5 meV in steps of 0.005 meV.}
    \label{fig:vanadiumNormalisationRealSpace}
\end{figure}

\begin{figure}[ht!]
    \centering
    \subfigure[]{\includegraphics[width=0.45\linewidth]{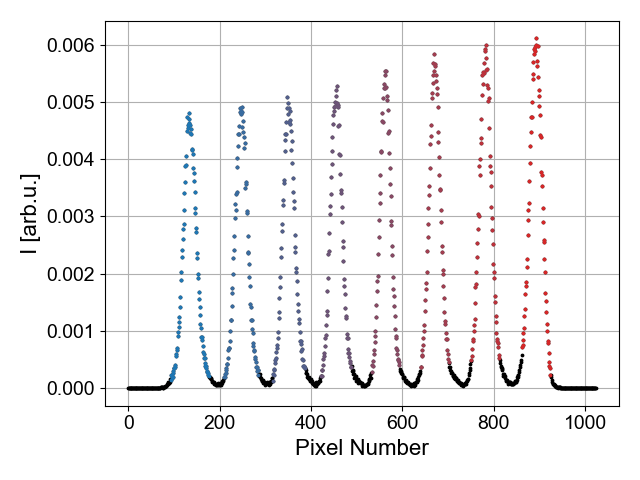}}
    \subfigure[]{\includegraphics[width=0.45\linewidth]{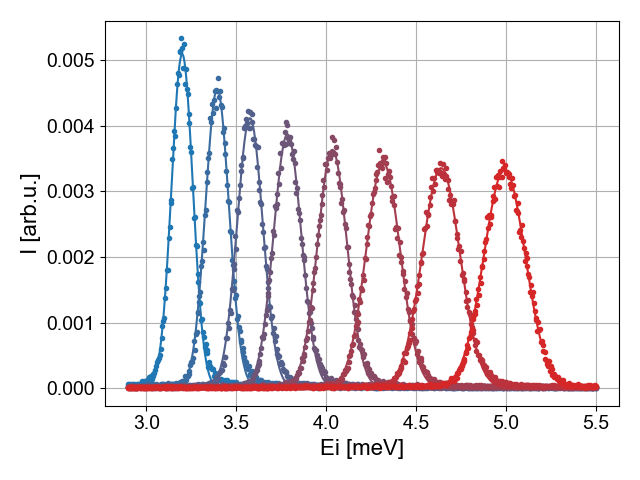}}
    \caption{(a) Energy integrated scattering intensity on a representative detector tube after an incident energy scan between 2.9 meV and 5.5 meV using a standard vanadium sample. The plot highlights the eight parts corresponding to the eight analyzers with the black dots signifying points outside the active area. (b) Scattering intensity in the eight analyzer-specific active areas along the detector tube as function of incoming energy. Note that the peaks in pixel space do not overlap. When converted to energy, the order of the peaks remains but but a quasi-continuous coverage is established.}
    \label{fig:vanadiumNormalisationEnergySpace}
\end{figure}

The final energy is defined by the eight analyzers segments and the prismatic concept. It can be determined by an extensive incident energy scan across all eight analyzer segments. Such a scan generates a data set as shown in Fig.~\ref{fig:vanadiumNormalisationRealSpace} for a hollow, cylindrical vanadium sample that is 1.8 cm in diameter and 1 cm tall. We use this material to cross-normalize the intensities between the different detector tubes. The intensity in each detector tube consists of eight Gaussian-shaped peaks corresponding to the eight analyzer segments (see Fig.~\ref{fig:vanadiumNormalisationEnergySpace}). Their fitted peaks (including data within $\pm$ 3$\sigma$ from the peak maximum) define the analyzer-specific active areas of the detector and correspondingly the final energies of the eight analyzer segments. In this case, the elastic energy resolution ranges from 145 to 278 $\mu$eV for E$_f$ = 3.2 to 4.987 meV (see Tab.~\ref{tab:PrismaticVanadiumMeanBinning1}). The final energies are consistent with the design\cite{Groitl2016}, however the energy resolution is somewhat larger than simulated for the secondary spectrometer alone. This results from the convolution of the primary and secondary spectrometer.

% C:\Users\lass_j\Documents\CAMEACommissioningPaper\VanadiumOverview.py

\begin{figure}[ht!] % Generated by: C:\Users\lass_j\Documents\CAMEACommissioningPaper\RealSimulationComparisons.py
    \centering
    \subfigure[Binning 1]{\includegraphics[width=0.45\linewidth]{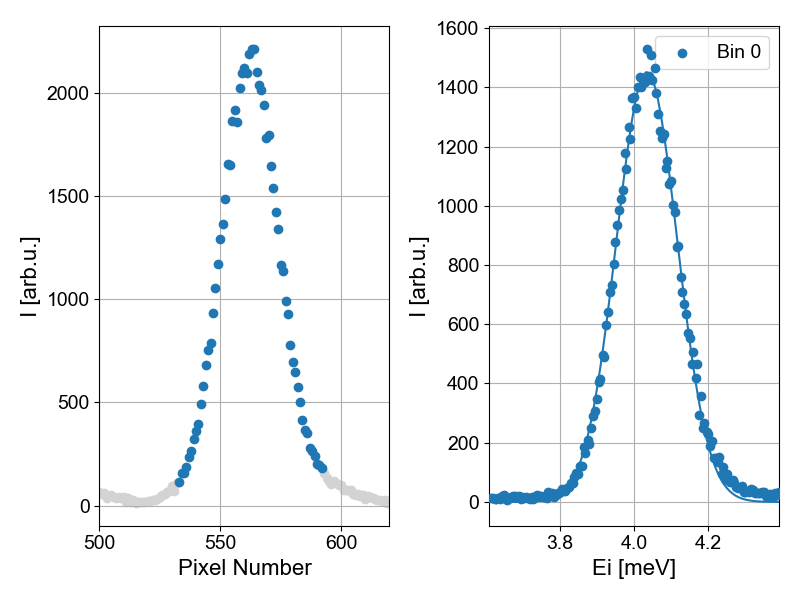}}
    \subfigure[Binning 3]{\includegraphics[width=0.45\linewidth]{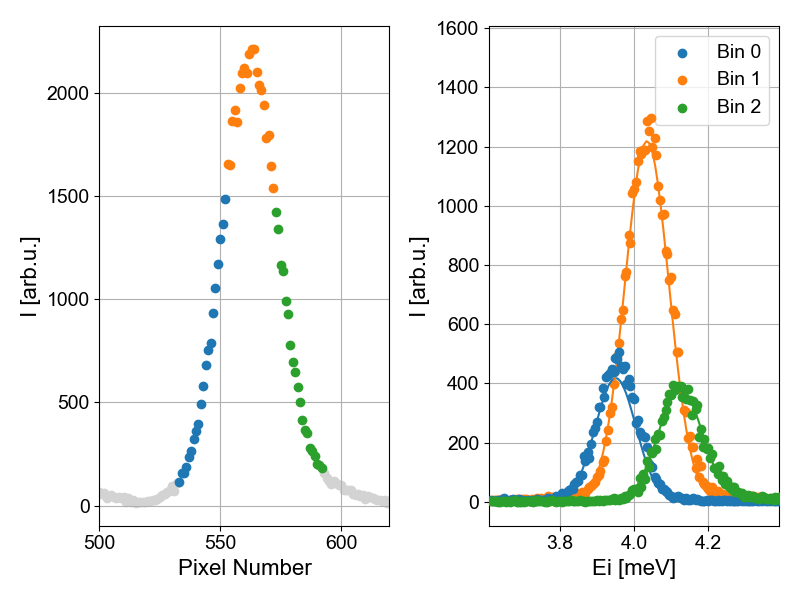}}
    \subfigure[Binning 5]{\includegraphics[width=0.45\linewidth]{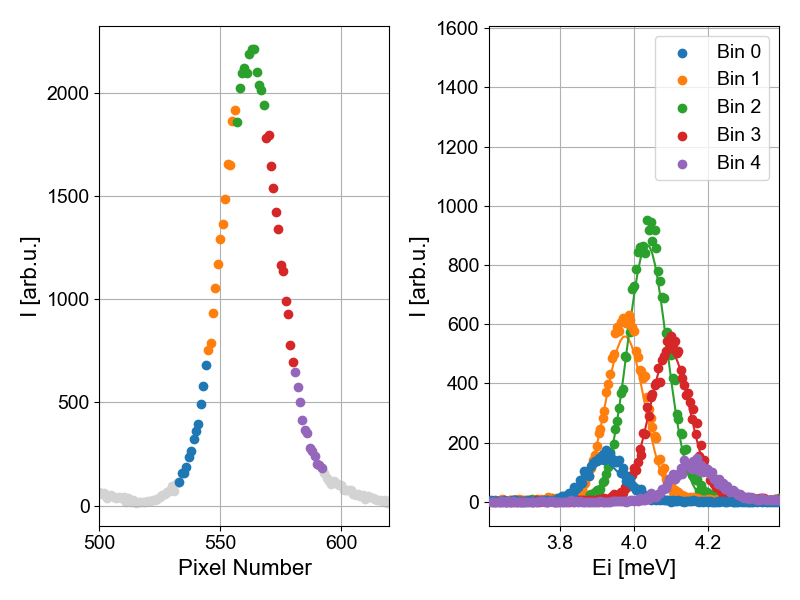}}
    \subfigure[Binning 7]{\includegraphics[width=0.45\linewidth]{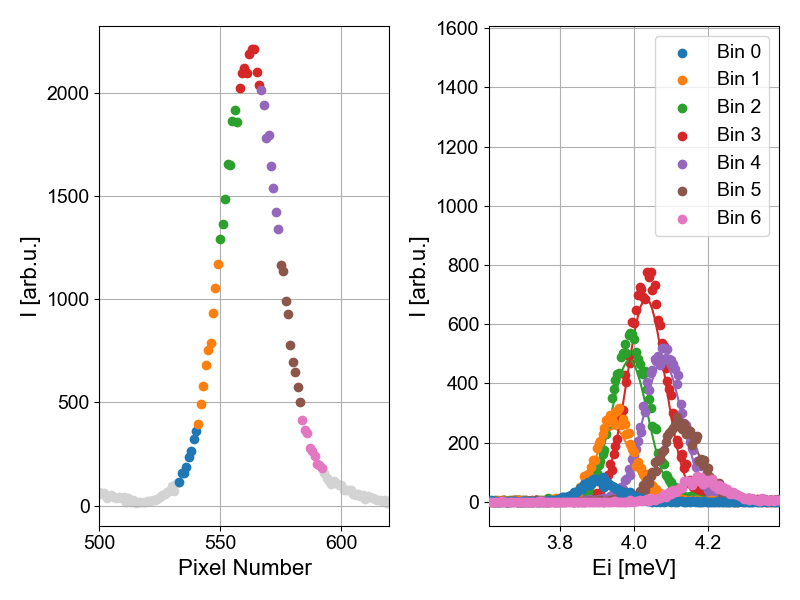}}
    \caption{Scattering intensity of the 4th analyzer segment and subdivision into 1, 3, 5 and 7 prismatic pixels for a representative detector tube shown in figures (a)-(d), respectively. On the left hand side the real space neutron count on the detector is shown while the right hand side shows the neutron count for the coloured, binned areas.}
    \label{fig:PristmaticTube41}
\end{figure}

Since the 8 analyzer blades are mounted on a Rowland sphere geometry with a focal point matching the distance between analyzer and detector, the prismatic concept allows one to further subdivide each analyzer segment into  smaller pixel arrays~\cite{Groitl2016}. Figure~\ref{fig:PristmaticTube41} shows the subdivision of the 4th analyzer ($E_f \sim $4.0 meV) into 1, 3, 5, and 7 prismatic pixels (for a representative detector tube). The associated final energies and elastic FWHMs are shown in Tab.~\ref{tab:PrismaticVanadiumanalyzer4Tube41}. We define the overall final energies and elastic resolution of CAMEA as the average over all detector tubes (see Tab.~\ref{tab:PrismaticVanadium}), and note that a subdivision into more than 8 pixels is inadvisable. We found that for finer binning the statistical uncertainty in defining individual Gaussians becomes exceedingly high compared to the additional improvement in energy resolution (see Tab.~\ref{tab:PrismaticVanadiumanalyzer4Tube41}). %The final energies and resolutions for different prismatic binnings from a representative detector tube are shown in Tab.~\ref{tab:PrismaticVanadiumanalyzer4Tube41}. 
The resolution of the prismatic pixellation has been simulated\cite{Groitl2016} but solely considering the secondary spectrometer. Here, resolutions comparable with binning 7 are found to range from 57.1 $\mu$eV at 3.21 meV to 116.5$\mu$eV at 5.01 meV. As for the non-prismatic simulation, these values are lower than those found at CAMEA due to the convolution with the primary spectrometer.

% Table generated using C:\Users\lass_j\Documents\Software\Alignment2022\BinningTable.py and normalization camea2022n000358
\begin{table}[ht!]
   \centering
    \begin{tabular}{|cc|ccccccc|}\hline 
    \multirow{3}{*}{Binning 1} &
      E$_f$ [meV] &   &   &   & 4.035 &   &   &  \\
      & FWHM [$\mu$eV] &   &   &   & 193 &   &   &  \\
      & Integrated Intensity [arb] &   &   &   & 746.9 &   &   &  \\\hline
    \multirow{3}{*}{Binning 3} &
      E$_f$ [meV] &   &   & 3.951 & 4.035 & 4.125 &   &  \\
      & FWHM [$\mu$eV] &   &   & 128 & 139 & 140 &   &  \\
      & Integrated Intensity [arb] &   &   & 143.1 & 451.3 & 142.5 &   &  \\\hline
    \multirow{3}{*}{Binning 5} &
      E$_f$ [meV] &   & 3.919 & 3.976 & 4.035 & 4.096 & 4.162 &  \\
      & FWHM [$\mu$eV] &   & 123 & 122 & 125 & 128 & 142 &  \\
      & Integrated Intensity [arb] &   & 42.9 & 182.0 & 288.8 & 174.6 & 46.7 &  \\\hline
    \multirow{3}{*}{Binning 7} &
      E$_f$ [meV] & 3.901 & 3.946 & 3.988 & 4.032 & 4.077 & 4.124 & 4.177\\
      & FWHM [$\mu$eV] & 118 & 115 & 118 & 121 & 123 & 126 & 146\\
      & Integrated Intensity [arb] & 18.8 & 76.0 & 150.3 & 221.1 & 155.5 & 84.1 & 27.3\\\hline
    \end{tabular}
    \caption{Average final energies, FWHMs, and normalization across all detectors as a function of different prismatic binnings for a representative detector tube and analyzer segment 4 ($E_f = ~$ 4.0 meV).}
    \label{tab:PrismaticVanadiumanalyzer4Tube41}
\end{table}

% Table generated using C:\Users\lass_j\Documents\Software\Alignment2022\BinningTable.py and normalization camea2022n000358
\begin{table}[H]
   \centering
    \begin{tabular}{|cc|cccccccc|}\hline 
    \multirow{3}{*}{Binning 1} &
      E$_f$ [meV] & 3.200 & 3.382 & 3.574 & 3.787 & 4.035 & 4.312 & 4.631 & 4.987\\
      & FWHM [$\mu$eV] & 145 & 154 & 165 & 182 & 196 & 218 & 247 & 278\\
      & Integrated Intensity [arb] & 541 & 599 & 694 & 771 & 832 & 876 & 964 & 999\\\hline
    \multirow{3}{*}{Binning 3} &
      E$_f$ [meV] & 3.200 & 3.383 & 3.574 & 3.786 & 4.035 & 4.311 & 4.630 & 4.984\\
      & FWHM [$\mu$eV] & 110 & 115 & 121 & 130 & 140 & 155 & 179 & 209\\
      & Integrated Intensity [arb] & 359 & 388 & 440 & 478 & 501 & 518.73 & 568 & 590\\\hline
    \multirow{3}{*}{Binning 5} &
      E$_f$ [meV] & 3.201 & 3.383 & 3.575 & 3.786 & 4.034 & 4.310 & 4.629 & 4.982\\
      & FWHM [$\mu$eV] & 98 & 102 & 107 & 115 & 126 & 140 & 163 & 191\\
      & Integrated Intensity [arb] & 236 & 251 & 284 & 304 & 323 & 336 & 365 & 377\\\hline
    \multirow{3}{*}{Binning 7} &
      E$_f$ [meV] & 3.202 & 3.383 & 3.575 & 3.786 & 4.034 & 4.310 & 4.628 & 4.982\\
      & FWHM [$\mu$eV] & 94 & 98 & 103 & 110 & 121 & 135 & 157 & 186\\
      & Integrated Intensity [arb] & 175 & 186 & 208 & 219 & 237 & 247 & 262 & 272\\\hline
    \multirow{2}{*}{RITA II} &
      Ef [meV] & - & - & 3.5 & - & 4.0 & - & 4.5 & 5.0 \\
      & FWHM [$\mu$eV] & - & - & 118 & - & 151 & - & 183 & 231 \\\hline
    \end{tabular}
    \caption{Average final energies, FWHMs, and normalization as a function of the central prismatic pixel and averaged over all detector tubes. Values for the replaced RITA II spectrometer are taken from \cite{RITAResolution}.}
    \label{tab:PrismaticVanadium}
\end{table}

% \section{Outlook}
\section{Experimental results and discussion}
\begin{figure}[H]
    \centering
    \subfigure[]{\includegraphics[width=0.47\linewidth]{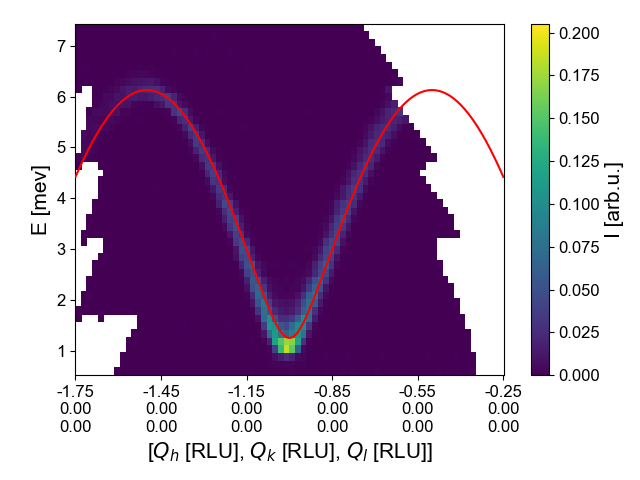}}
    \subfigure[]{\includegraphics[width=0.47\linewidth]{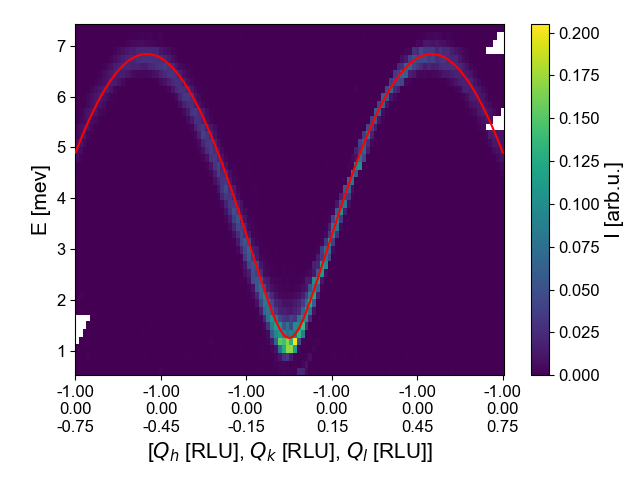}}
    \caption{Spin-wave dispersion of MnF$_2$ measured on CAMEA and over-plotted with the fitted spin-wave model using the parameters in Tab.~\ref{table:MnF2}. The cut in (a) is along ($H$, 0, 0), while in (b) along the (-1, 0, $L$) in reciprocal lattice units (rlu).}
    \label{fig:MnF2Plot}
\end{figure}

The final part of the commissioning consisted of a standard experiment, measuring the spin waves and determining the exchange parameters of a MnF$_2$ single crystal. The 6.2 g large sample was aligned in the horizontal ($H$, 0, $L$)-plane. The spin waves of the spin 5/2 compound have been studied in the past~\cite{Okazaki1964, Yamani2010}. This allows us to efficiently check the instrument's performance over a sizeable excitation range CAMEA is covering. The bandwidth of the magnetic spin waves in MnF$_2$ extend from roughly 1 to 7.2 meV. Since the analyzer arrays of CAMEA cover an energy range of 1.8 meV, we used four different incoming energies to map out the spin-wave spectrum of MnF$_2$. Two detector tank positions shifted by 4$^\circ$ were used per incident energy to cover all dark spots of the modular analyzer design. For each $E_i$-$2\theta$ combination we performed a sample rotation scan of 150 degrees in steps of 2 degrees. At every rotation angle we counted at a monitor rate of 200k, corresponding to a counting time of 48 s at $E_I$ = 5 meV and 75 s at 11.5 meV, respectively. A subset of this 8.4 hours long measurement is displayed in Fig.~\ref{fig:MnF2Plot}, where we show scattered intensity along the two major crystallographic directions in the plane as a function of energy transfer. The data were converted and visualized with our software package MJOLNIR~\cite{Lass2020} using a prismatic binning of 8. The data quality, spin-wave continuity and their intensity homogeneity confirm a successful commissioning of the spectrometer. The two-dimensional cuts along the two principal crystallographic directions and as a function of energy transfer further reveal the resolution of the spectrometer, as it is known that the spin wave of MnF$_2$ is intrinsically resolution limited~\cite{Yamani2010}. In particular we point out that the focus and de-focus sides of the instrument resolution function are clearly visible in the magnon dispersion around (-1, 0, 0) in reciprocal lattice units (rlu) (see Fig.~\ref{fig:MnF2Plot}). 

The dispersion of the antiferromagnetic spinwave of MnF$_2$ can be calculated analytically~\cite{Yamani2010, Okazaki1964}, enabling a direct refinement of the magnetic exchange parameters from the results shown in Fig.~\ref{fig:MnF2Plot}. We obtained them by fitting the spin-wave over a 41 x 41 grid spanning from -1.5 $<H<$ -0.5 rlu and -1 $<L<$ 0.5 rlu. At each point we performed an energy cut with a step size of 0.05 meV and integrated over a circular shape of diameter 0.02 \AA\ in momentum transfer. The 1681 data points were then refined against the microscopic Hamiltonian describing the spin-wave dispersion of MnF$_2$, and we tabulated them in Tab.~\ref{table:MnF2}. Here, $J_1$ and $J_2$ describe the Heisenberg interactions between identical and different Mn sites, respectively, and $D_{d-d}$ accounts for the single ion anisotropy in the system. Our findings were compared to reported values of Okazaki \textit{et. al.}~\cite{Okazaki1964} (see Tab.~\ref{table:MnF2}). While the overall agreement with the reported results is satisfying, we find small disagreements in $J_1$ and $D_{d-d}$. We note that similar discrepancies have also been reported by Yamani \textit{et. al.}~\cite{Yamani2010}, also showcasing that the fitting parameters slightly vary for different cutting directions through the dispersion as a result of the resolution ellipsoid of the employed triple-axis spectrometer. As such the massively enhanced density of fitted excitation energies along multiple different directions in our CAMEA experiment leads to a more accurate results of the interaction parameters in MnF$_2$ when compared to these previous experiments, and testifies a successful commissioning of the CAMEA spectrometer at PSI. 

% C:\Users\lass_j\Documents\Software\Scripts\ResolutionCAMEA\MnF2Fitting.py
\begin{table}[H]
    \centering
    \begin{tabular}{|c|c|c|c|} \hline
     &   $J_1$ [meV] & $J_2$ [meV] & $D_{d-d}$ [meV] \\ \hline
    Current Findings &   0.0354(6)  & 0.1499(3) & 0.131(6) \\ \hline
    Earlier measurements\cite{Okazaki1964} & 0.028 & 0.152 & 0.091 \\ \hline
    \end{tabular}
    \caption{Magnetic exchange parameters of MnF$_2$ refined from a measurement at CAMEA and compared to earlier results reported in Ref.~\cite{Okazaki1964}.}
    \label{table:MnF2}
\end{table}

We note that a meaningful performance comparison between a CAMEA-like instrument and a standard triple-axis instrument is problematic. At every sample rotation angle, 6656 points are measured simultaneously at CAMEA (using a prismatic binning of 8), which is contrasted by the single point that is measured on a triple-axis spectrometer at an identical configuration. However, while in a triple-axis experiment every point is chosen on purpose, a large point density over an energy range of 1.8 meV and 60$^\circ$ scattering angle window is measured on CAMEA. As such, an experiment on a CAMEA-like instrument generates data that resemble that from ToF spectrometers. However, a direct comparison to this instrument type is also challenging. This is because the energy bandwidth of CAMEA is limited to 1.8 meV and has an out-of-plane acceptance of $\pm$ 2$^{\circ}$ only. Thus, a direct comparison between the different instrument types is sensible only in specific instrument configurations where the energy resolution between the spectrometers are very different and only if the out-of-plane coverage of the ToF instrument is omitted. This, however, also underlines the complementary between the different types of neutron spectrometers, allowing the user to choose the optimal instrument for their scientific problem.

\subsection{Summary and Outlook}
In summary we report on the successful commissioning of the cold neutron multiplexing secondary spectrometer CAMEA at SINQ, PSI, including the neutron guide and its double-focusing monochromator. The instrument allows for an efficient mapping of horizontally scattered low-energy neutrons under extreme conditions down to $T$ = 30 mK that can be combined with magnetic fields up to $\mu_0H$ = 15 T and/or high pressure options. The multiplexing detector system enables a quasi-continuous coverage over an energy bandwidth of 1.8 meV and an angular range of 61$^\circ$ with an elastic energy resolution down to $\sim$0.1 meV FWHM in a single sample rotation scan. Combined with the instrument's high flux and low background the instrument constitutes a complementary extension of state-of-the-art neutron time-of-flight and triple-axis spectrometers. CAMEA will enable new insight into persistent questions in the field of quantum matter and may open new routes for future condensed matter research. Furthermore, the insights we have gained at CAMEA will be valuable for all upcoming spectrometers employing the CAMEA principle. This, for instance, includes the BIFROST spectrometer at the ESS.
  
Finally we comment on several options to further improve the performance of the spectrometer. Valuable hardware extensions may involve the implementation of a velocity selector, which would further suppress the background of the instrument and reduce accidental scattering arising from the monochromator or the analyzer crystals. A polarization option would allow for an efficient discrimination between nuclear and magnetic signals, and would enable an efficient distinction between longitudinal, transverse, and chiral magnons. Finally, we note that a plethora of improvements can be achieved via digital competences. Promising efforts in this direction may include exact calculations of the resolution ellipsoid, non-parametric background models and the implementation of machine learning algorithms into the data analysis.

\section*{Acknowledgements}
We thank Simon Ward for his assistance regarding the use of the SpinW. We further thank Kim Lefmann and Rasmus Toft-Petersen for fruitful discussions. The financial support of the Swiss National Science Foundation (R'Equip 206021\_144972) is gratefully acknowledged. Jakob Lass was supported by the Danish National Committee for Research Infrastructure through DanScatt Grant No 7129-00006B. H.J. was supported by the EU Horizon 2020 program under the Marie Sklodowska-Curie Grant No 701647, and the Carlsberg Foundation. 

\section*{AUTHOR DECLARATIONS}

\section*{Conflict of Interest}

The authors declare no competing interests.

\section*{DATA AVAILABILITY}

 All data were generated at the Paul Scherrer Institut and are available from the corresponding authors upon reasonable request.
 
\bibliography{main}
\end{document}